\documentclass{article}
\usepackage{amsfonts,amsmath}
\mathchardef\mhyphen="2D % Define a "math hyphen"
\usepackage[a4paper]{geometry}
\usepackage{microtype}
\usepackage[shortlabels]{enumitem} %package lets us do enumeration by roman
\usepackage{relsize}

\def\zo/{$0\mkern2mu\mhyphen1$}%0-1

\def\nn/{$n \times n$}

\title{Four Amazing Positivities with Dimers/$i$-Matchings}
\author{Paul Federbush\\
Department of Mathematics\\
University of Michigan\\
Ann Arbor, MI, 48109-1043\\
pfed@umich.edu}
\usepackage{amsthm} % Needed for conj*

\newtheorem{conj*}{Conjecture} % Defines non-numbered conjecture environment

\numberwithin{lemma}{section}
\numberwithin{conj*}{section}
\begin{document}

\maketitle
\begin{abstract}
	We collect a number of striking recent results in a study of dimers on infinite r-regular bipartite lattices and also on r-regular bipartite graphs. We clearly separate rigorously proven results from conjectures. A primary goal is to show people: here is a field which is ripe for further interesting research. We separate four classes of endeavor, of which we here extract two items to whet one's appetite. Primo, the  hyper-rectangular lattices of every dimension have their first 20 virial coefficients  positive. (One has no understanding of this yet!) Secondo, all regular bipartite graphs with less than $14$ vertices satisfy graph positivity, defined below. (Here there is some understanding.)
\end{abstract}
	Much of this work was joint with P. Butera and M. Pernici. Most of the computer work was by M. Pernici. As said above, we have four sections, in each we separate the proved portion from the conjectures. Notes at the end present some further material. Proofs may be computer based, analytic, or analytic with computer assistance. (All computations are in integer arithmetic.)

	In each section we italicize the conjecture in it towards which the effort is centered. The general references are for I, \cite{1}, for II, \cite{3}, for III, \cite{2} and for IV, \cite{3}. Additional references are separately noted.

%%%%%% FIRST SECTION %%%%%%%%%%%%%

\section*{Section I}

\subsection*{I\,.\,Proved}\label{I-proved}
\begin{enumerate}
	\item For an infinite $r$-regular lattice the dimer entropy, $\lambda$, as a function of the dimer density, $p$, for $p$ small enough is of the form
		\begin{equation}
	\lambda(p) = \frac{1}{2}(p \ln(r) - p\ln(p) - 2(1-p)\ln(1-p)-p) + \sum_{k=2}^\infty a_k p^k.
\end{equation}
This result was first obtained by me \cite{6}. It is an easy consequence of previously known results.

\item \label{IP-2} For the infinite hyper-rectangular lattices of every dimension the $a_k(d)$ are positive for $k \leq 20$. (See Note 2.)

\item Note 1 below lists other infinite lattices for which all computed $a_k$ are positive.
\end{enumerate}

\subsection*{I\,.\,Conjectured}\label{I-conjectured}
	\begin{enumerate}[resume]
	\item {\itshape For an infinite r-regular vertex-transitive bipartite lattice all $a_k$ are positive. Results in [9], [10], and [11] strongly
argue for the restriction to vertex-transitive lattices we here have
added, a modification of our original conjecture.}
\end{enumerate}

%%%%%%%%%%% SECOND SECTION %%%%%%%%%%%%

\section*{Section II}

\subsection*{II\,.\,Proved}
\begin{enumerate}
	\item  For a dimer gas on an infinite hyper-rectangular lattice of any dimension the first 20 virial coefficients are positive. ( See Note 2. )
	\item Note 1 lists other infinite lattices for which all computed virial coefficients are positive (for a dimer gas).
\end{enumerate}

\subsection*{II\,.\,Conjectured}

	\begin{enumerate}[resume]
	\item {\itshape For a dimer gas on an infinite r-regular bipartite vertex-transitive lattice, all virial coefficients are positive. }
	
\end{enumerate}

%%%%%%%%%%%%%%% THIRD SECTION %%%%%%%%%%%%%%%

\section*{Section III}
\subsubsection*{Definitions}
We deal with $r$-regular bipartite graphs with $v=2n$ vertices. We define $d(i)$ a function of the number of $i$-matchings, $m_i$
\begin{equation*}
	d(i)\equiv \ln\left(\frac{m_i}{r^i}\right)-\ln\left(\frac{\overline{m}_i}{(v-1)^i}\right)
\end{equation*}
where $\overline{m}_i$ is the number of $i$-matchings on the complete (not bipartite complete) graph on the same vertices.
\begin{equation*}
	\overline{m}_i = \dfrac{v!}{(v-2i)!i!2^i}.
\end{equation*}
We let $\Delta$ be the finite difference operator
\begin{equation*}
	\Delta d(i) = d(i+1) -d(i).
\end{equation*}
We say a graph satisfies graph positivity if all the meaningful $\Delta^k d(i) \geq 0$. (Loosely speaking, the conditions $\Delta^k d(i) \geq 0$ are a generalization of the positivity of the $a_k$ in eq.(1) from infinite lattices to general graphs.)

\subsection*{III\,.\,Proved}
\begin{enumerate}
	\item All graphs with $v < 14$ satisfy graph positivity.
	\item When $r=4$ the first violations occur when $v=22$, in 2 graphs of $2806490$. 
	\item For $r=3$ the fraction of graphs not satisfying graph positivity monotonically decreases between $v=14$ and $v=30$. There is a single violation at $v=14$. 
	\item If $r \leq 10$, $i+k \leq 100$, $k\leq 27$, or
$i+k<30$ all $r$, then
		\begin{equation*}
			\mathrm{Prob}(\Delta^k d(i) \geq 0)~ {\xrightarrow[n\to \infty]{} 1}
		\end{equation*}
		This is proved in \cite{4}. (See Note 2.)
\end{enumerate}
\subsection*{III\,.\,Conjectured}
{\itshape 
	\begin{enumerate}[resume]
		\item As $n$ goes to infinity the fraction of graphs that satisfy graph positivity approaches one.

\end{enumerate}

}

\section*{Section IV}

\subsubsection*{Definitions}
Again, dealing with $r$-regular bipartite graphs with $v=2n$ vertices. $m_i$ the number of $i$-matchings. We define 
\begin{equation*}
	u_i = -\ln(i!m_i).
\end{equation*}
We say a graph satisfies virial positivity if $\Delta^k u(i)\geq 0$ for all meaningful $k \geq 2$.
(Loosely speaking a generalization of an upper bound on the virial coefficients from infinite lattices to all graphs.)

\subsection*{IV\,.\,Proved}
\begin{enumerate}
	\item All graphs with $v \leq 18$ satisfy virial positivity.
	\item When $r = 4$, the first violation occurs when $v=20$ in a single graph among 62611 graphs with $v=20$.
	\item When $r=3$ the fraction of graphs not satisfying virial positivity decreases monotonically between 18 and 30 with a single violation at 18. 
	\item If $r \leq 10$, $i+k \leq 100$, $2 \leq k\leq 27$, or
$i+k<30$ and all $r$,  then 
		\begin{equation*}
			\mathrm{Prob}(\Delta^k u(i) \geq 0) \xrightarrow[n \to \infty]{} 1.
		\end{equation*}
		This is proved in \cite{5}. (See Note 2.)
\end{enumerate}

\subsection*{IV\,.\,Conjectured}
{\itshape 
	\begin{enumerate}[resume]
		\item As $n \to \infty$ the fraction of graphs that satisfy virial positivity approaches one.

\end{enumerate}

}

%%%%%%%%%%% NOTES %%%%%%%%%%%%%%

\section*{Note 1}\label{note-1} 
Additional regular bipartite lattices for which computed coefficients are positive
\begin{enumerate}[a)]
	\item Hexagonal lattice, through $7^{\text{th}}$ coefficient.
	\item Tetrahedral lattice, through $19^{\text{th}}$ coefficient.
	\item Hyper-body-centered-cubic lattices of dimension $d=3,4,5,6,7$ through $24^{\text{th}}$ coefficient.
\end{enumerate}
These results are given in \cite{1} with other references there cited.

\section*{Note 2}\label{note-2}

In Sections I and II the rigorous proof that for all $d$ the first 20
terms are positive  arises from [9] and [8], without these one would only be able to prove the first 10 coefficients were positive. Beyond the results implicit in [4], in Sections
III and IV the extension to region $i+k<30$ and all $r$, arises from
the computations reported in [12].

\end{document}